\documentclass[%
 reprint,
 amsmath,amssymb,
 aps,onecolumn
]{revtex4}

\usepackage{natbib}
\usepackage{aas_macros}
\usepackage{graphicx}
\usepackage{dcolumn}
\usepackage{bm}

\newcommand\ba{\begin{eqnarray}}
\newcommand\ea{\end{eqnarray}}

\usepackage{color}

\begin{document}


\title{21cm forest probes on the axion dark matter in the post-inflationary Peccei-Quinn symmetry breaking scenarios}

\author{Hayato Shimabukuro}
 \affiliation{Yunnan University, SWIFAR,No.2 North Green Lake Road, Kunming, Yunnan Province,650500,China}
 \email{shimabukuro@ynu.edu.cn}
\author{Kiyotomo Ichiki}
 \affiliation{%
Graduate School of Science, Division of Particle and Astrophysical Science, Nagoya University, Chikusa-ku, Nagoya, 464-8602, Japan
}%
 \affiliation{%
 Kobayashi-Maskawa Institute for the Origin of Particles and the Universe, Nagoya University, Chikusa-ku, Nagoya, 464-8602, Japan
 }%
  \email{ichiki.kiyotomo@c.mbox.nagoya-u.ac.jp}
\author{Kenji Kadota}
\affiliation{
 Center for Theoretical Physics of the Universe, Institute for Basic Science (IBS), Daejeon, 34051, Korea
}
 \email{kadota@ibs.re.kr}

\date{\today}

\begin{abstract}
We study the future prospects of the 21cm forest observations on the axion-like dark matter when the spontaneous breaking of the global Peccei-Quinn (PQ) symmetry occurs after the inflation. The large isocurvature perturbations of order unity sourced from axion-like particles can result in the enhancement of minihalo formation, and the subsequent hierarchical structure formation can affect the minihalo abundance whose masses can exceed ${\cal O}(10^4) M_{\odot}$ relevant for the 21cm forest observations. We show that the 21cm forest observations are capable of probing the axion-like particle mass in the range $10^{-18}\lesssim m_a \lesssim 10^{-12}$ eV for the temperature independent axion mass. For the temperature dependent axion mass, the zero temperature axion mass scale for which the 21cm forest measurements can be affected is extended further to as big as of order $10^{-6}$ eV.
\end{abstract}

\maketitle

\section{Introduction}

There can arise ubiquitous light degrees of freedom, such as the pseudo-Goldstone bosons as a consequence of the spontaneous symmetry breaking of an approximate symmetry in the early Universe, and the cosmic perturbations arising from those light fields are of great interest as the target for the cosmological observations. A typical example is the axion from the breaking of the Peccei-Quinn (PQ) symmetry to address the QCD strong-CP problem which can also serve as a promising dark matter candidate \cite{Peccei:1977hh,Weinberg:1977ma,Wilczek:1977pj,Preskill:1982cy,Abbott:1982af,Dine:1982ah,sik1983,raf1987,as2009,kad2015,ana2017,Kadota:2013iya,kel2017,hua2018,hook2018,kad2015b,har1992,fed2019,kad2019,shi2019}. If the PQ symmetry breaks after the inflation (or if the PQ symmetry is restored during the reheating epoch and broken later), so-called post-inflationary PQ symmetry breaking scenarios, the axions can lead to the isocurvature fluctuations of order unity and result in the enhancement of the structure formation at the small scales \cite{hog1988,kolb1994,zu2006,Ringwald:2015dsf,marsh15,sh1994,tk1998,2017PhRvL.119c1302I,Feix:2019lpo,Feix:2020txt}. 

We study how much the small halo abundance is affected by these axion isocurvature perturbations and, as one of the promising tools to study the small scale structures, we estimate the 21cm forest signal due to the hyperfine structure of neutral hydrogen atoms in the small halos. The 21cm forest is a system of absorption lines appearing in the spectra by radio background sources due to intervening neutral hydrogen atoms in analogy to the Lyman-$\alpha$ forest, while the 21cm forest can explore much smaller scales ($k \gtrsim 10{\rm Mpc}^{-1}$) than the Lyman-$\alpha$ forest \cite{2002ApJ...577...22C,2002ApJ...579....1F,2006MNRAS.370.1867F, Chabanier:2018rga,2015aska.confE...6C}. We focus in particular on the halos whose virial temperature is less than $10^4{\rm K}$ for which the atomic cooling is ineffective, so that we can expect the abundant neutral hydrogen atoms due to the insufficient star formation in those small halos.

We find that 21cm forest can distinguish the axion isocurvature model from the conventional adiabatic perturbation model (without the isocurvature modes) for the axion mass $4\times 10^{-18} \lesssim m_a \lesssim 2 \times 10^{-12}$ when the axion mass is temperature independent. We also discuss the axion parameters when the mass is temperature dependent, for which the axion mass at the zero temperature sensitive to the 21cm forest observations extends to an even higher mass range.

In Section \ref{sec:pq}, we start with a brief review on how the large (of order unity) isocurvature fluctuations arise in the post-inflation PQ symmetry breaking scenarios. We then discuss the 21cm forest signals in existence of such axion isocurvature perubations in Section \ref{sec:halo_gas}. Section \ref{sec:ax} finds the concrete axion model parameters which can affect the observable 21cm absorption lines, followed by the discussion/conclusion section.

\section{Post-inflation PQ symmetry breaking scenario}
\label{sec:pq}
We consider the scenarios where the PQ symmetry breaking occurs after the inflation, so that we can expect the large isocurvature perturbations produced during the radiation domination epoch \cite{hog1988,kolb1994,zu2006}.
When the global U(1) symmetry is spontaneously broken for the potential $V(\phi)=\lambda (|\phi|^2 -f_a^2/2)^2$, the complex PQ field $\phi$ settles down at the the minimum $\phi=(f_a/\sqrt{2}) e^{i \theta}$. We identify the axion as the angular field $a\equiv f_a \theta$ with an axion decay constant $f_a$ which sets the PQ symmetry breaking scale. For the post-inflation PQ-breaking scenarios, because each horizon patch is causally disconnected, the axion vacuum expectation values among different horizon volumes can be different depending on randomly distributed angles $\theta \in [-\pi,\pi]$. The axion can acquire the mass $m_a$ from the non-perturbative effects inducing the axion potential $m_a^2 a^2/2$, and the axion oscillation starts when the Hubble scale $H(t)$ becomes comparable to the potential curvature $m_a=3H$. The initial axion field amplitude and the resultant energy density in each causally disconnected Hubble path are different, and one can expect large axion density fluctuations over the different horizons.
We need specify the initial power spectrum in estimating the evolution of axion dark matter density fluctuations, which we set to be the power spectrum when the axion stars oscillation \cite{fai2017,2020AJ....159...49D}. 
The total power spectrum for the matter over-density is the sum of the standard adiabatic power spectrum (presumably induced by the inflation) and the axion isocurvature power spectrum $P_{{\rm iso}}$.
The axion isocurvature fluctuations are smoothed inside the horizon scale because of the gradient term in the Lagrangian (Kibble mechanism \cite{kib1976}) and are uncorrelated white noise due to the randomly distributed angles beyond the horizon scale, so that we consider the initial isocurvature modeled as the white noise power spectrum with the sharp-$k$ cut-off 
\begin{equation}
P_{\mathrm{iso}}(k,t_{\mathrm{osc}})= P_0\Theta(k_{\mathrm{osc}}-k),P_0=\frac{24}{5}\frac{\pi^2}{k_{\mathrm{osc}}^3}
\end{equation}
where $\Theta$ is the Heaviside function and $k_{{\rm osc}}$ is the comoving wave number when the axion starts oscillation. The normalization factor $P_0$ is obtained using the average values of uniformly distributed angle. More concretely, we estimate the initial axion density fluctuation of order unity $ \langle \delta_a^2 \rangle=4/5 $ ( $\delta_a \equiv (\rho_a-\bar{\rho}_a)/ \bar{\rho}_a$)). We here used the axion density $\rho_a\propto \theta^2$ and the uniformly distributed angle such that $\langle \theta^2 \rangle =\pi^2/3, \langle \theta^4 \rangle =\pi^4/5$. The normalization factor $P_0$ is consequently obtained from the relation for the variance $\sigma^2 \equiv \langle \delta_a^2 \rangle = (2\pi)^{-3} \int P(k) d^3 k$. We illustrate our findings assuming the axion constitutes the whole dark matter of the Universe unless stated otherwise.
The total power spectrum for the matter over-density is the sum of standard adiabatic power spectrum and the axion isocurvature power spectrum \cite{fai2017,2020AJ....159...49D}
\begin{equation}
P(k,z)=P_{\mathrm{ad}}(k) D^2(z)+\Theta(k_{\mathrm{osc}}-k)\frac{24\pi^2}{5k_{\mathrm{osc}}^3}\left(\frac{D(z)}{D(z_{*})}\right)^2\left(\frac{1+z_{\mathrm{eq}}}{1+z_{*}}\right)^2
\label{totpower}
\end{equation}
\begin{figure}[htbp]
\includegraphics[width=0.8\hsize]{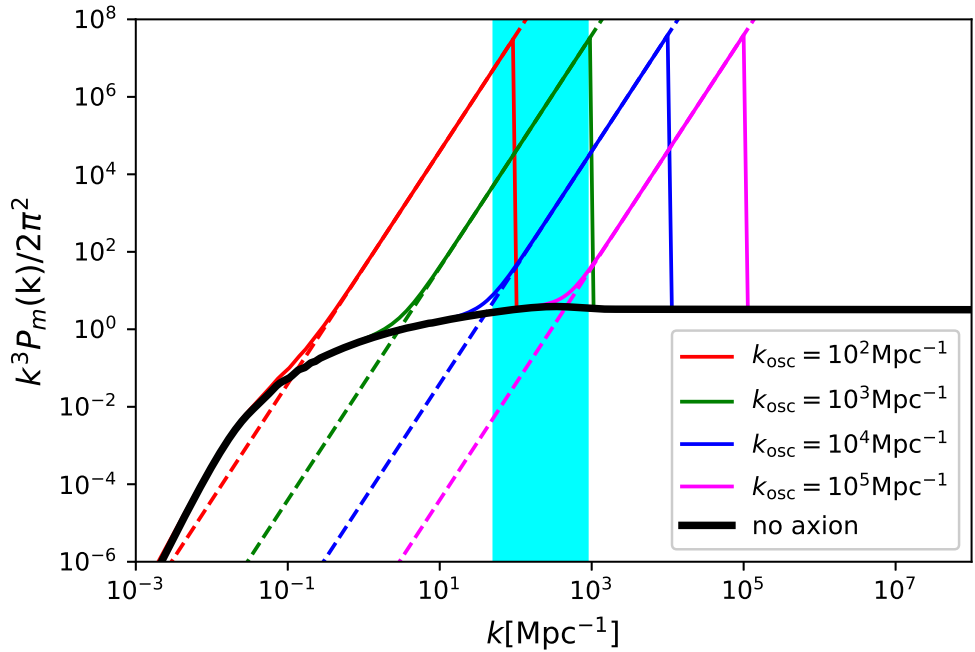}
\caption{Matter power spectra at $z$=10 including the axion isocurvature modes to be compared with the adiabatic CDM power spectrum without the isocurvature modes. We vary $k_{{\rm osc}}=,10^2$(red), $10^3$ (green),$10^4$ (blue),$10^5$ (magenta) $\mathrm{Mpc}^{-1}$. The shaded region represents the wavenumber scales corresponding to mass scales determined by minimum and maximum masses the 21cm forest observations are sensitive to at $z=10$.}
\label{fig:ps}
\end{figure}
where $z_*$ is an arbitrarily chosen redshift deep in the matter domination epoch and $z_{{\rm eq}}\sim 3400$ is the matter-radiation equality epoch. 
$D$ represents the growth factor and Fig. \ref{fig:ps} shows the corresponding matter power spectra at $z=10$ (the reference redshift value for the estimation of the 21cm signals in our study) along with the conventional adiabatic CDM power spectrum without the axion isocurvature modes. We can see the small scale power is indeed enhanced by the addition of the isocurvature modes, without affecting the large scale power where the adiabatic modes dominate the isocurvature modes. We can hence infer that the small halo abundance can be enhanced in existence of the axion isocurvature perturbations. From Fig.\ref{fig:ps}, we can expect that $k_{\mathrm{osc}}\sim 10^2-10^5$ ${\rm Mpc}^{-1}$ affects the minihalo formation and thus the number of 21cm absorption lines. More precise estimation using the analytical modeling of the 21cm forest signals will be the main subject of the following section.

The 21cm forest is indeed not sensitive to the large halos where the hydrogen atoms are ionized. For concreteness, the maximum mass in our studies is chosen to be the halo mass corresponding to $T_{{\rm vir}}=10^4$ K below which the gas atomic cooling through the atomic transition is inefficient for the star formation to keep the hydrogen atoms neutral \cite{2001PhR...349..125B} (we assume the molecular hydrogen cooling is negligible because the molecular hydrogen is expected to be photo-dissociated by Lyman-Werner radiation backgrounds emitted by stars \citep[e.g.][]{2001ApJ...548..509M,2007ApJ...671.1559W})
\ba
M_{\mathrm{max}}(z)
=
3.95 \times 10^7
\left(
\frac{\Omega_m h^2 }{ 0.15}
\right)^{-1/2}
\left(
\frac{1+z}{10}
\right)^{-3/2}
M_{\odot}
\label{eq:mmax1}
\ea
The minimum mass $M_{{\rm min}}$, on the other hand, is chosen to be the baryon Jeans mass$^1$ \footnotetext[1]{Note the mass range to estimate the 21cm forest signals could be effected by the non-trivial gas heating processes (see, for instance, \citep[]{1998MNRAS.296...44G,2011MNRAS.418..906T} for the time-averaged filtering mass taking account of the time evolution of gas to respond to earlier heating). }
\ba
M_{min}(z)=
5.7 \times 10^3
\left(
\frac{\Omega_m h^2}{0.15}
\right)^{-1}
\left(
\frac{\Omega_b h^2}{0.02}
\right)^{-3/5}
\left(
\frac{1+z}{10}
\right)^{3/2}
M_{\odot}
\label{eq:M_min}
\ea
Fig. \ref{fig:ps} also shows the range indicating the comoving wavenumber scales relevant for our 21cm signal estimation 
($50\mathrm{Mpc} \lesssim k_{\mathrm{osc}} \lesssim 900\mathrm{Mpc}$ corresponding to $M_{\mathrm{min}}=7.7 \times 10^3 M_{\odot}<M_{\mathrm{halo}} < M_{\mathrm{max}}=3.8 \times 10^7 M_{\odot}$ at $z=10$.).
We can see that a relatively large $k$ value is relevant for the 21cm forest observations, and the 21cm signals are calculated by considering the contributions of the minihalos covering those relevant halo masses.

We first study how the number of 21cm absorption lines are affected by the isocurvature fluctuations based on the power spectrum Eq. \ref{totpower} treating $k_{osc}$ as a free parameter. Our aim here is to find what range of $k_{osc}$ the future 21cm signals can be sensitive to. We then discuss the corresponding axion mass range which the 21cm forest observations can probe by specifying the axion models.

\section{21cm forest observations}

\subsection{Mass function}
\label{massf}

\begin{figure}[htbp]
\includegraphics[width=0.7\hsize]{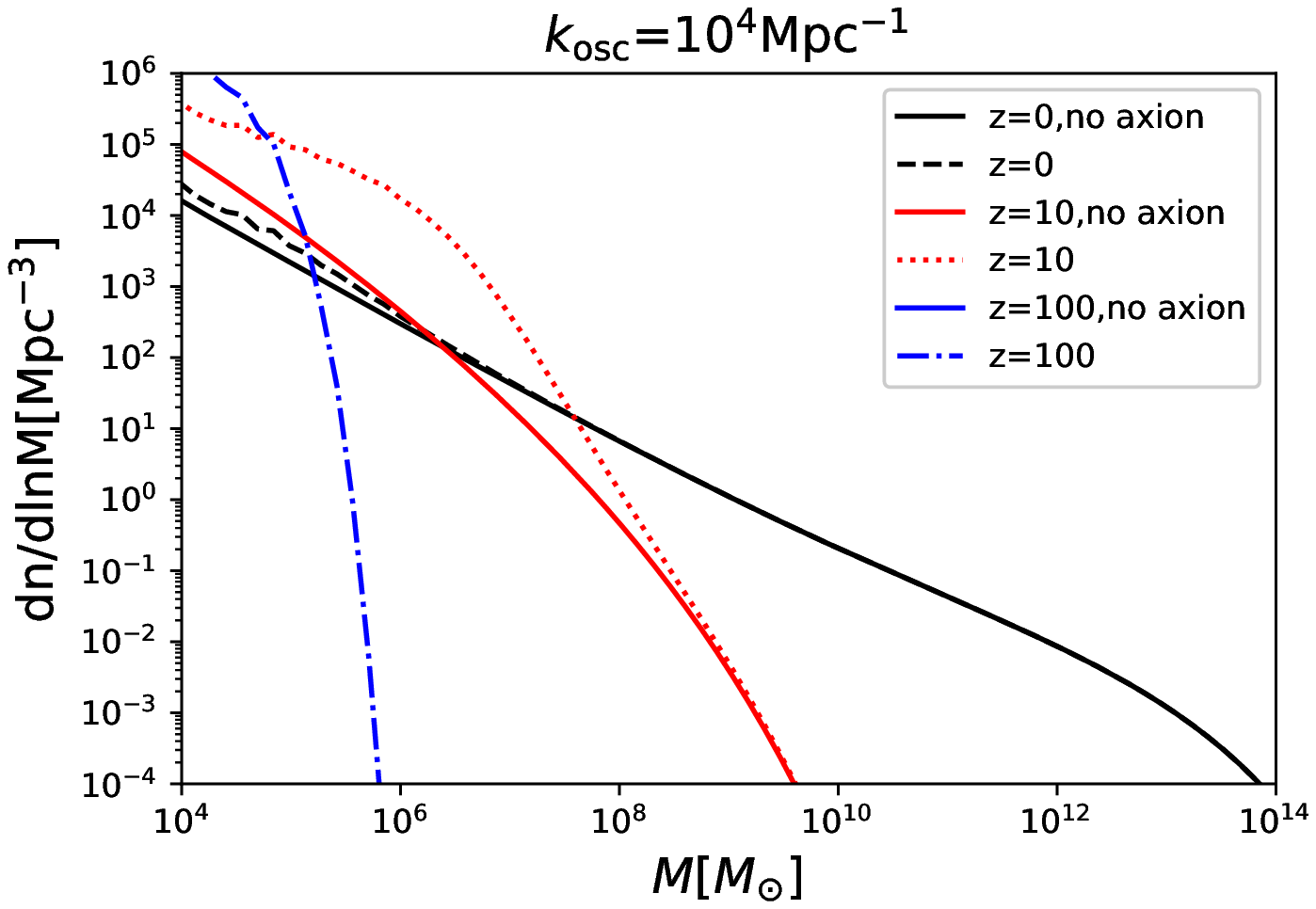}
\includegraphics[width=0.7\hsize]{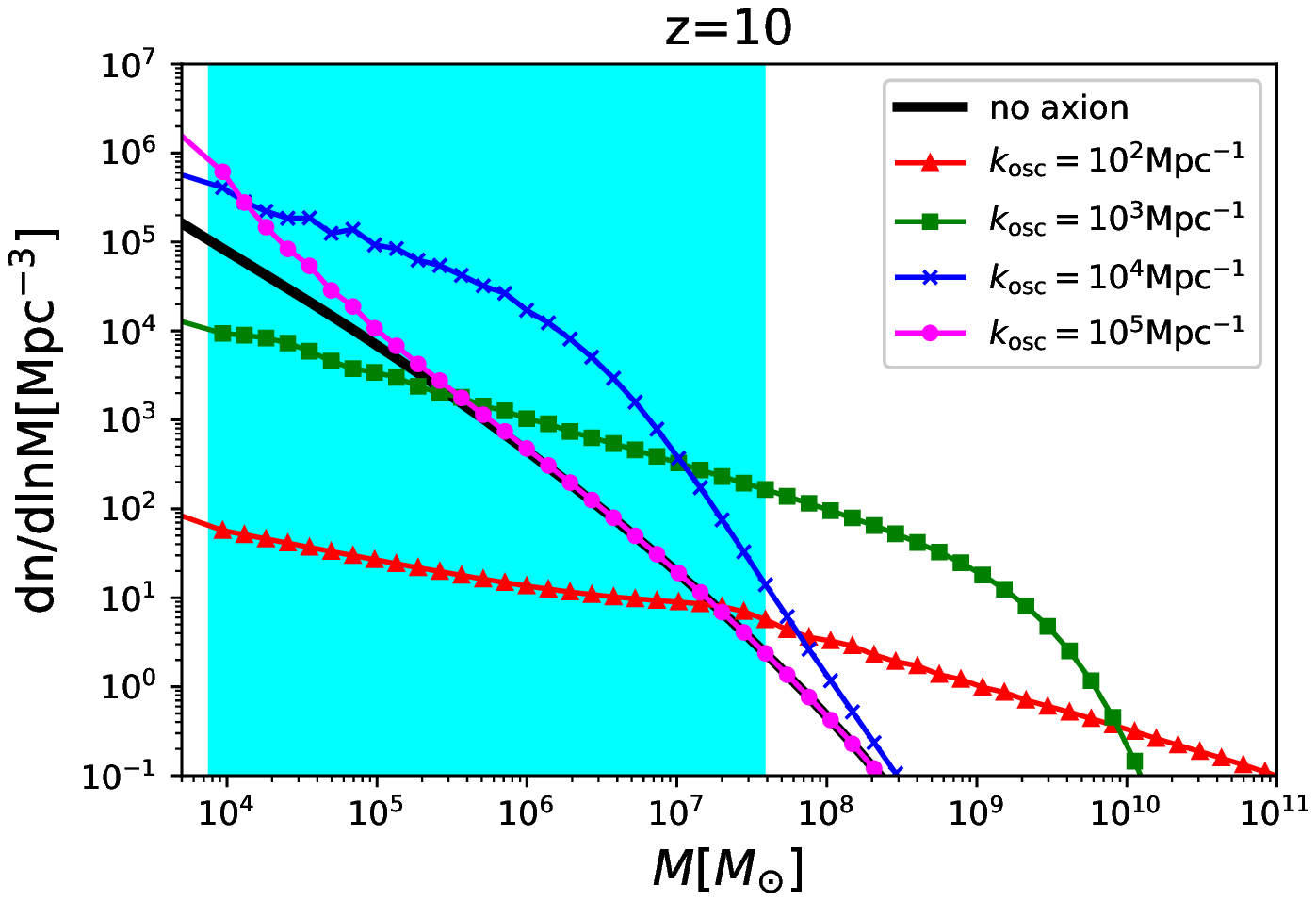}
\caption{({\it Top}) The mass functions including axion isocurvature fluctuations at z=0({\it dashed}), 10({\it dotted}), 100({\it dot-dashed}) for $k_{\mathrm{osc}}=10^{4} \mathrm{Mpc}^{-1}$. For comparison we also show no axion case({\it solid lines}). At $z=100$, the mass function is too small to be shown for no axion. ({\it Bottom}) The halo mass function at 
  $z=10$ for $k_{{\rm osc}}=10^2\mathrm{Mpc}^{-1} ({\it triangle}),10^{3}\mathrm{Mpc}^{-1} ({\it square}),10^4\mathrm{Mpc}^{-1} ({\it cross}),10^5 \mathrm{Mpc}^{-1}({\it circle})$ and no axion ({\it solid}) (the last two are almost identical in this figure). The shaded region($7.7\times 10^3 M_{\odot} \lesssim M\lesssim 3.8\times 10^{7} M_{\odot}$) represents the mass range the 21cm forest observations can probe.
}
\label{fig:MF_z}
\end{figure}

Before performing the 21 cm signal calculations from the minihalos, let us first illustrate how the large axion dark matter fluctuations can help enhance the abundance of the {\it large} minihalos with the mass range relevant for the 21cm forest observations.
The top panel of Fig.\ref{fig:MF_z} shows the proper halo number density and illustrates a significant abundance of minihalos at $z= 100$ while it is negligible for the conventional adiabatic curvature perturbation scenario at such a high redshift \cite{1974ApJ...187..425P}. These minihalos are then assembled to form the larger halos in the hierarchical structure formation.
At a reference redshift of $z=10$ for our 21cm forest observations, we can indeed see the abundance of minihalos larger than that of the conventional adiabatic CDM scenarios without the axion isocurvature modes. This hierarchical structure formation continues until the minihalos have merged into the larger halos whose formation is dominated by the large adiabatic fluctuation contributions (consequently the abundance difference disappears), even though we can still see a slight difference in the small halo abundance even at $z=0$ in this figure. The lower panel of Fig. \ref{fig:MF_z} shows the mass function at our reference redshift $z=10$ for different values of $k_{\mathrm{osc}}$, where we also showed the mass range relevant for the 21cm forest observations (bounded by $M_{\mathrm{min}}$ and $M_{\mathrm{max}}$ given in Eqs. \ref{eq:mmax1} and \ref{eq:M_min}). A smaller $k_{{\rm osc}}$ has a larger isocurvature amplitude $\propto k_{\mathrm{osc}}^{-3}$, and the small halos can start forming earlier which also results in the earlier formation of bigger halos due to the merging of small halos. Consequently, for a relatively small value of $k_{{\rm osc}}$ such as $k_{{\rm osc}}=100~\mathrm{Mpc}^{-1}$ in this figure, the halo abundance can be smaller (bigger) for a small (large) halo mass than that of the CDM scenario whose halo formation epoch is delayed compared with the axion isocurvature scenario. Too small a value of $k_{\mathrm{osc}}$, say of order ${\cal O}(100)\mathrm{Mpc}^{-1}$ as shown in this figure, hence would be disadvantage for our purpose (even though the isocurvature mode amplitude is big) because the halo abundance enhancement shows up in the mass range insensitive to the 21cm forest observations and indeed the minihalo abundance is suppressed for the minihalo mass range of our interest.

Note these axion dark matter minihalos relevant for the 21cm forest observations are far larger than the first gravitationally collapsed axion minihalos which can span a wide range ${\cal O}(10^{-10\sim 2}) M_{\odot}$ depending on the axion models (around $10^{-10} M_{\mathrm{sun}}$ for the QCD axion) \cite{hog1988,kolb1994,zu2006,har2016,ena2017,fai2017,vis2018,va2018,2020AJ....159...49D}. Those {\it small} minihalos are however not expected to affect our discussions because the 21cm forest observations are not sensitive to such small halos, and we restrict our discussions to the {\it large} minihalos whose masses exceed the baryon Jeans mass ($\gtrsim {\cal O}(10^4) M_{\odot}$). 


\subsection {21cm forest signals}\label{sec:halo_gas}

The 21cm forest, which is in analogy to the Lyman alpha forest, is a system of 21cm absorption lines that appear in the continuum spectrum of luminous background sources\cite{2002ApJ...577...22C,2002ApJ...579....1F,2006MNRAS.370.1867F}. The 21cm forest can be produced by a neutral hydrogen atom in the IGM,  sheet and filament structures in the cosmic web and the collapsed objects called minihalos. Among them, the column density in the minihalos is larger and hence produces larger optical depth than the others, resulting in the strong 21cm absorption lines. We hence focus on the 21cm forest by neutral hydrogen gas in minihalos in our work.
We briefly outline how we analytically estimate the 21cm forest signals by minihalos, following \cite{2002ApJ...579....1F,2014PhRvD..90h3003S,shi2019}.

One first needs to specify the dark matter halo and gas profiles for the analytical estimations of the 21cm signals  \cite{2002ApJ...579....1F, 2001PhR...349..125B}. We assume the NFW profile \cite{1997ApJ...490..493N,2000ApJ...540...39A} for the dark matter density distribution inside the virial radius
\begin{equation}
r_{{\rm vir}}=0.784\bigg(\frac{M}{10^{8}h^{-1}M_{\odot}}\bigg)^{1/3}\bigg[\frac{\Omega_{m}}{\Omega_{m}^{z}}
\frac{\Delta_{c}} {18\pi^{2}}\bigg]^{-1/3} \bigg(\frac{1+{\it z}}{10}\bigg)^{-1}h^{-1}[{\rm kpc}]
\label{eq:virial_radius}
\end{equation}
where $\Delta_{c}=18\pi ^{2}+82d-39d^{2}$ is the halo over-density collapsing at a redshift ${\it z}$, $d=\Omega_{m}^{z}-1$ and
$\Omega_{m}^{z}=\Omega_{m}(1+z)^{3}/(\Omega_{m}(1+z)^{3}+\Omega_{\Lambda})$. We further assume the concentration parameter ${\it y}=r_{{\rm vir}}/r_{{\rm s}}$ ($r_s$ is the scale radius) scales as $(1+z)^{-1}$ following the N-body simulation results for the halos at a high redshift \cite{2005MNRAS.363..379G}. Given the NFW dark matter density profile, one can obtain the analytical solution for the gas density assuming the isothermal profile in the hydrostatic equilibrium \cite{1998ApJ...497..555M,2011MNRAS.410.2025X}
\begin{equation}
\ln \rho_{{\rm g}}(r)=\ln \rho_{{\rm g0}}-\frac{\mu m_{{\rm p}}}{2k_{{\rm B}}T_{{\rm vir}}}[v_{{\rm esc}}^{2}(0)-v_{{\rm esc}}^{2}(r)],
\label{eq:gas_profile1}
\end{equation}
where $\mu=1.22$ is the mean molecular weight of the gas and $m_{{\rm p}}$ is the
proton mass.
The central gas density $\rho_{{\rm g0}}$ is normalized by the cosmic value of $\Omega_{b}/\Omega_{m}$ and given by
\begin{equation}
\rho_{g0}(z) =
\frac{(\Delta_c/3)y^{3}e^{A}}{\int_{0}^{y}(1+t)^{A/t}t^{2}dt} \left(\frac{\Omega_b}{\Omega_m}\right)\bar{\rho}_{m}(z)~,
\end{equation}
where $A=3y/F(y)$ and $\bar{\rho}_{m}(z)$ is the mean total matter density at a redshift $z$.
The virial temperature reads
\begin{equation}
  T_{{\rm vir}} =1.98\times 10^{4}\bigg(\frac{\mu}{0.6}\bigg)\bigg(\frac{M}{10^{8}h^{-1}M_{\odot}}\bigg)^{2/3}\bigg[\frac{\Omega_{m}}{\Omega_{m}^{z}}\frac{\Delta_{c}}{18\pi^{2}}\bigg]^{1/3}\bigg(\frac{1+z}{10}\bigg)[{\rm K}]
\end{equation}
and the escape velocity is
\begin{equation}
  v_{{\rm esc}}^{2}(r) = 2\int_{r}^{\infty}\frac{GM(r^{'})}{r^{'2}}dr^{'}=2V_{c}^{2}\frac{F(yx)+yx/(1+yx)}{xF(y)}~,
\label{eq:esc_velocity}
\end{equation}
where $x\equiv r/r_{{\rm vir}}$ and $F(y)=\ln(1+y)-y/(1+y)$ with the circular velocity 
\begin{equation}
  V_{c}^{2} =  \frac{GM}{r_{{\rm vir}}}=23.4\bigg(\frac{M}{10^{8}h^{-1}M_{\odot}}\bigg)^{1/3}
  \bigg[\frac{\Omega_{m}}{\Omega_{m}^{z}}\frac{\Delta_{c}}{18\pi^{2}}\bigg]^{1/6}\bigg(\frac{1+z}{10}\bigg)^{1/2}[{\rm km/s}].
  \label{eq:cir_velocity}
\end{equation}
We adopt this gas density profile for the neutral hydrogen gas in a minihalo. The 21cm forest signals have been discussed in details in the literature, and we only give here the relevant equations (see for instance \cite{2002ApJ...579....1F,2006PhR...433..181F,1997ApJ...475..429M,2014PhRvD..90h3003S,shi2019} and references therein for the derivations).
The photons emitted from the radio loud sources at a high redshift are absorbed by the intervening neutral hydrogen gas in the minihalos. The corresponding optical depth experienced by a photon going through a minihalo is given by \cite{2002ApJ...579....1F}
\begin{equation}
  \tau(\nu,M,\alpha)=\frac{3h_{{\rm p}}c^{3}A_{10}}{32\pi k_{{\rm B}}\nu_{21}^{2}}
  \int_{-R_{{\rm max}}(\alpha)}^{R_{{\rm max}}(\alpha)}dR\frac{n_{{\rm H{\sc I}}}(r)}{T_{{\rm S}}(r)\sqrt{\pi}b}\exp\bigg(-\frac{v^{2}(\nu)}{b^{2}}\bigg),
\label{eq:optical1}
\end{equation}
where $r^2=\alpha^2+R^2$, $\alpha$ is an impact parameter, $R_{{\rm max}}$ is the maximum radius of the halo at $\alpha$, $\nu$ represents the frequency of a photon at emission from the source ($(1+z)/(1+z_{source})=\nu_{21cm}/\nu$), $M$ is a halo mass and $A$ is the Einstein coefficient for the spontaneous transition. The exponential factor represents the Doppler broadening with $v(\nu)=c(\nu-\nu_{21})/\nu_{21}$ and the velocity dispersion $b^2(r)=2kT_{vir}/m_p$ (we assume the gas kinetic temperature equals the virial temperature, which would be a reasonable simplification because the gas cooling is assumed to be inefficient in a minihalo). $T_S$ is the spin temperature in a minihalo \cite{2005ApJ...622.1356Z,2006MNRAS.370.1867F,2014PhRvD..90h3003S}, and it approaches the virial temperature of the minihalo in the inner region and the CMB temperature in the outer part because the gas density is small there(see Fig.1 in \cite{2014PhRvD..90h3003S}).

Based on our modeling for a single halo, we can now estimate the density of 21cm absorption lines from the multiple halos in the observed spectrum per redshift as
\begin{equation}
  \frac{dN(>\tau)}{dz}=\frac{dr}{dz}\int_{M_{{\rm min}}}^{M_{{\rm max}}}dM\frac{dN}{dM}\pi r^{2}_{\tau}(M,\tau), 
  \label{eq:abundance1}
\end{equation}
where $dN/dM$ is the halo mass function and $\pi r^{2}_{\tau}$ represents the cross section of a halo with the impact parameter $r_{\tau}$ with the optical depth exceeding $\tau$ \cite{2006MNRAS.370.1867F}.
There can be several choices for the mass function form, and the mass function parameters at a high redshift of our interest can well be different from the conventionally adopted values for those at a low redshift. We regardless simply use the Press-Schechter mass function \cite{1974ApJ...187..425P} in the following discussions, which would suffice for our purpose of seeing the difference from the conventional cosmological models without the axions and illustrating the potential power of the 21cm forest observables on the axion parameters (we checked using the Sheth-Tormen mass function \cite{1999MNRAS.308..119S} gives a small difference and does not affect our discussions).

\begin{figure}[htbp]
\includegraphics[width=1.0\hsize]{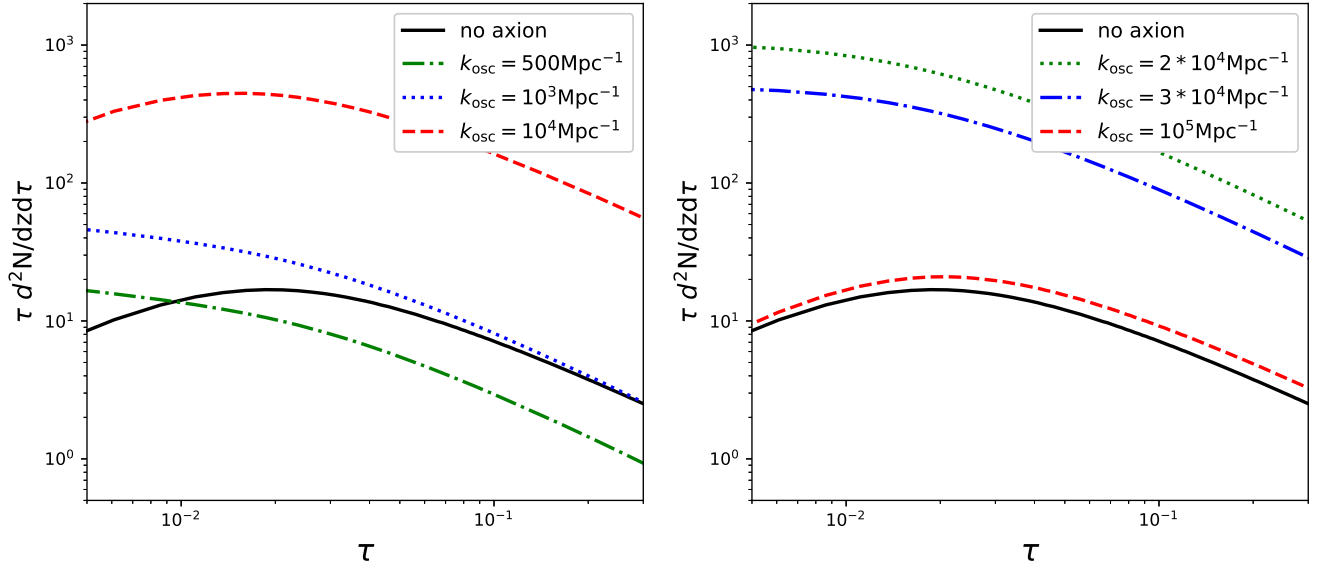}
\caption{The number of 21cm absorption lines as function of optical depth $\tau$ for various $k_{osc}$.}
\label{fig:21cm_abs}
\end{figure}

The actual observations can see the absorption lines only for a sufficiently large optical depth, and we calculate the number of absorption lines $n^{21}$ with a minimum optical depth $\tau_{\mathrm{min}}$
\ba
n^{21}=\int _{\tau_{\mathrm{min}}} d\tau \int dz  \frac{d^2N}{d\tau dz}
\ea
The absorption line abundance as a function of the optical depth is illustrated in Fig. \ref{fig:21cm_abs}.
This figure confirms our discussions in \S \ref{massf} that a small $k_{\rm osc}$ ($k_{\rm osc}=500{\rm Mpc^{-1}}$ in this figure) would lead to the suppression of the signals compared with the conventional adiabatic case and the larger $k_{\rm osc}$ can enhance the abundance of the minihalos relevant for the 21cm forest signals. Too large a value of $k_{{\rm osc}}$ however makes the isocurvature amplitude $\propto k_{\rm osc}^{-3}$ too small to affect the 21cm forest observations and we can find there is an optimal value $k_{\rm osc}\sim 2\times 10^4{\rm Mpc}^{-1}$ which makes the signals biggest.
To find the value of $k_{\rm osc}$ which can make the 21cm forest signals distinguishable from the adiabatic CDM scenario, we consider the condition
$n^{21}_{\rm CDM}+\Delta n^{21}_{\rm CDM}<n^{21}_{k_{\rm osc}}-\Delta n^{21}_{k_{\rm osc}} $ with $\Delta$ representing the corresponding error. For our order of magnitude estimation of $k_{\mathrm{osc}}$ of our interest, we assume the number of absorption lines obey the Poisson statistics with $\Delta n^{21}=\sqrt{n^{21}}$ as a 1-$\sigma$ requirement to find the relevant $k_{{\rm osc}}$. The value of $\tau_{{\rm min}}$ depends on the sensitivity of the experiment, and we use, for concreteness, $\tau_{\rm min}=0.03$ and assume the redshift bin width $\Delta z=1$ in our estimate. As a result, we numerically obtain 
$530{\rm Mpc}^{-1} <k_{{\rm osc}}<4.8\times 10^5{\rm Mpc}^{-1}$ 
so that our axion scenarios can be distinguished from the pure adiabatic scenarios using the 21cm forest observations.
We now study what axion models can realize these desirable comoving scales $k_{\rm osc}$.

\section{Axion parameters}
\label{sec:ax}
Now that we find the range of $k_{\mathrm{osc}}$ the 21cm observations are sensitive to, we discuss what axion parameters, especially the axion mass, are the potential targets of the 21cm forest measurements by considering the concrete axion models.
The axion acquires the mass from the non-perturbative effects and the axion oscillation starts when
\ba
m_a(T_{\mathrm{osc}})=3 H(T_{\mathrm{osc}})
\ea
The mass dependence on the temperature is conventionally parameterized as
\ba
m_a = m_{0} \left( \frac{T}{\mu} \right)^{-n}
\ea
$\mu$ represents the (possibly hidden sector) strong coupling scale and we for concreteness parameterize it as $\mu=\sqrt{m_{a,0} f_a}$ in our discussions and in the figures $^2$ \footnotetext[2]{The precise expressions for $\mu$ is model dependent. For instance, for the QCD axion, $\mu \sim \Lambda_{QCD} \sim 2.5 \sqrt{m_a f_a} \sim 200$ MeV \cite{har2016,fai2017}}. The axion mass is temperature independent for $T<\mu$, $m_a(T<\mu)\equiv m_{a,0}$.

The index $n$ can control how quickly the mass can switch on. While $n=4$ for the conventional QCD dilute instanton gas model, $n$ can take different values based on a model and we simply treat it as a free parameter$^3$ \footnotetext[3]{See for instance Ref. \cite{har2016,fai2017,va2018} discussing the axion minicluster properties for $n=0$ up to $n=20$. The lattice QCD simulations and interacting instanton liquid model, for example, give the slightly smaller values than $4$ and some non-QCD axion-like particle models can give $n=0$\cite{wan2009,dia2014,bo2016}.}. The Hubble value at the radiation-matter equality is of order $H(T_{\mathrm{eq}})\sim 10^{-28}${\rm eV}, and we consider the ultra-light particle mass range $m_a \gg 10^{-27} {\rm eV} \gtrsim 3 H(T_{\mathrm{eq}})$ for which the scalar field starts oscillations during the radiation domination epoch.
  
\begin{figure}[htb!]
  \begin{center}
    \begin{tabular}{cc}
      
      \includegraphics[width=1.0\hsize]{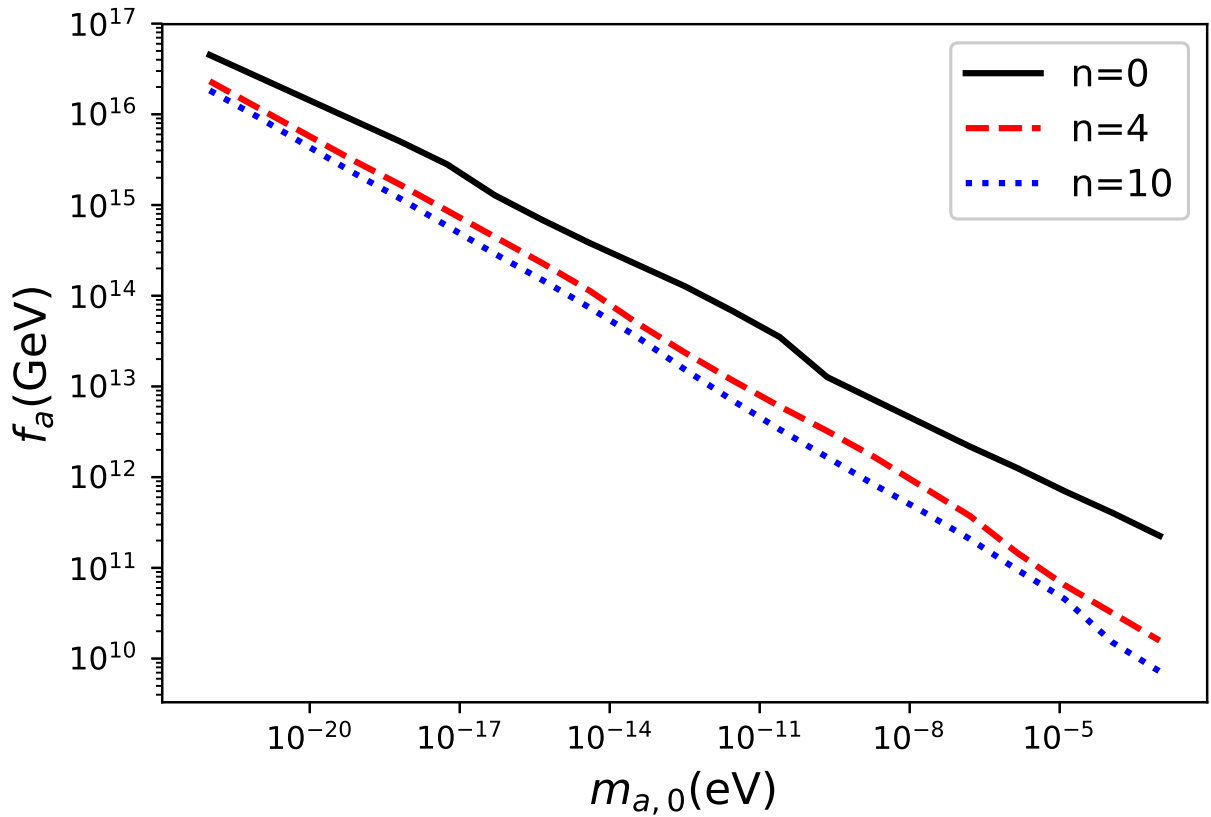}  
      
    \end{tabular}
 
  \end{center}
  \caption{
  The value of the axion decay constant $f_a$ to satisfy $\Omega_a h^2=0.12$ as a function of the zero temperature axion mass $m_{a,0}$.
}
\label{figkjkosc}
\end{figure}

\begin{figure}[htb!]
  \begin{center}
    \begin{tabular}{cc}
    \includegraphics[width=1.0\hsize]{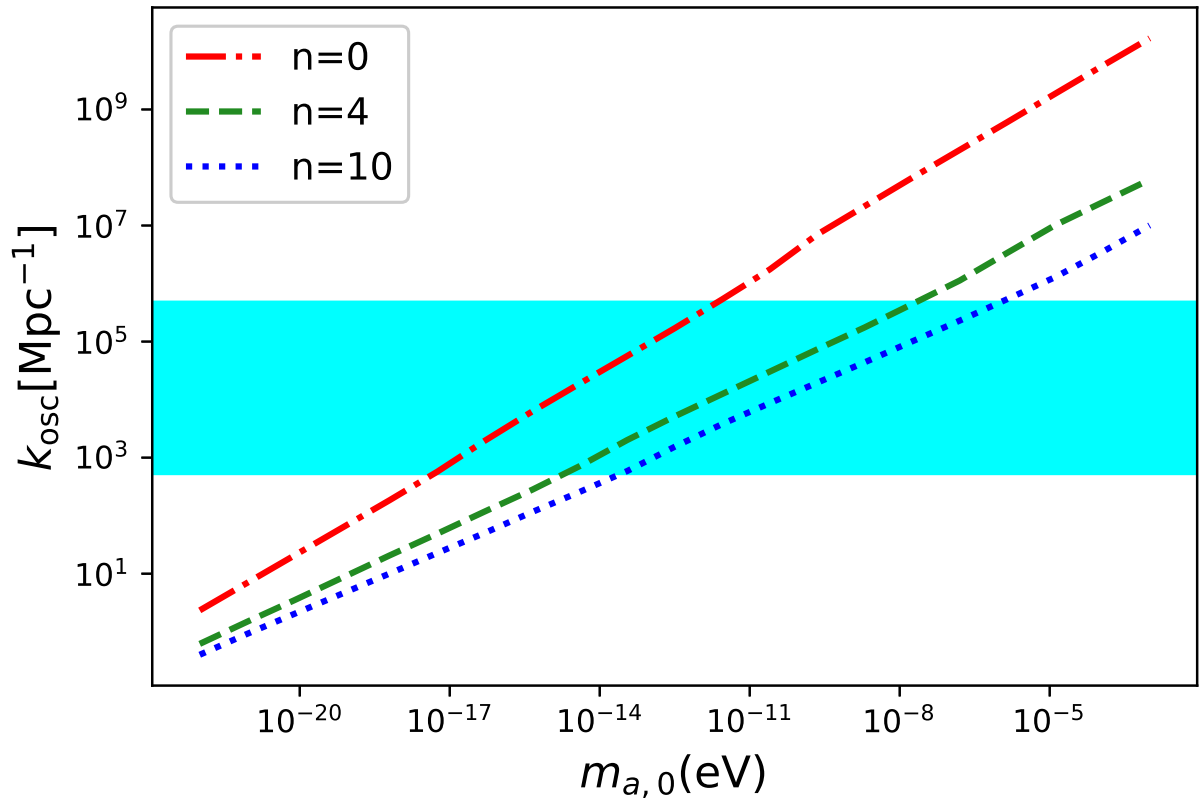}
    \end{tabular}
 
  \end{center}
  \caption{The comoving horizon scale when the axion oscillation initiates as a function of $m_{a,0}$ for the different temperature dependence of axion mass (represented by different values of $n$). The shaded region represents the scales that 21cm forest can probe.  
}
\label{figkjkosc2}
\end{figure}

We have illustrated our discussions assuming the post-inflation PQ symmetry breaking scenarios where the axions make up the whole cold dark matter of the Universe (the analogous discussions can be applied to the partial dark matter scenarios, where the isocurvature amplitude is modified to be proportional to $(\Omega_a/\Omega_{\mathrm{CDM}})^2$). This can hence fix the value of $f_a$ for a given $m_{a,0}$ for consistency, and the values of the axion decay constant satisfying $\Omega_a h^2=0.12$ as a function of the zero temperature axion mass $m_{a,0}$ are shown in Fig. \ref{figkjkosc} for reference (note not all the axion mass range in this figure can be probed by the 21cm forest observations to be shown below). 
 For this figure, the current cold dark matter axion density is estimated by, noting that the axion number density scales as $n_a \propto R^{-3}$ ($R$ is a scale factor) once it starts oscillation behaving as the matter, 
\ba
\rho_a(T_{\mathrm{now}})=m_{a,0} \frac{\rho_a(T_{\mathrm{osc}})}{m_a(T_{\mathrm{osc}})}
\left(
\frac{R(T_{\mathrm{osc}})}{R(T_{\mathrm{now}})}
\right)^3
\ea
where $\rho_a(T_{osc})=m_a(T_{osc})^2 f_a^2 \theta^2$/2 ($ \langle \theta^2 \rangle =\pi^2/3$ for the randomly distributed angle $\theta \in [-\pi,\pi]$).
Fig. \ref{figkjkosc2} shows, as a function of $m_{a,0}$, the corresponding comoving horizon scale $k_{\mathrm{osc}}$ when the axion starts oscillation $k_{\mathrm{osc}}=R(T_{\mathrm{osc}})H(T_{\mathrm{osc}})$. The effective relativistic degrees of freedom $g_*$ in Ref. \cite{hus2016} is used in our analysis.
In this figure, we also superimposed the range of $k_{\mathrm{osc}}$ we found in the last section to indicate the range for which the 21cm forest observations can distinguish between the axion models and the pure adiabatic model without the axion isocurvature perturbations.
Table.\ref{table:mass} lists the minimum and maximum axion mass which can be explored by the 21cm forest for $n$=0,4,10.
For the temperature independent axion mass ($n=0$), the axion mass parameter range the 21cm forest can probe is $4 \times 10^{-18} \mathrm{eV}\lesssim m_{a,0} \lesssim 1.8 \times 10^{-12} \mathrm{eV} $).
It is interesting that the 21cm forest measurements can be sensitive up to the mass range $m_{a,0}\sim 10^{-12}$ eV, which goes well beyond and is complementary to the Lyman-$\alpha$ observations which can currently put the lower bound $m_{a,0} \gtrsim 10^{-17}$ eV \cite{Irsic:2019iff}. The bounds from the observables on the larger scales such as the future CMB and BAO can also be complimentary, which is expected to give the bounds of order $m_a\gtrsim 10^{-18}$ eV \cite{Feix:2019lpo,Feix:2020txt}. For a bigger temperature dependence, say for $n=10$, the zero temperature axion mass sensitive to the 21cm forest observations can even become as big as $m_{a,0}\sim 10^{-6}$ eV. 

\begin{table}[htbp]
\scalebox{1.5}[1.5]{
  \begin{tabular}{l|c|rr}
     & $m_{\mathrm{a,0,min}}[\mathrm{eV}]$ & $m_{\mathrm{a,0,max}} [\mathrm{eV}]$ \\ \hline
     
    $n=0$ & $4 \times 10^{-18}$  & $1.8\times 10^{-12}$   \\  
    
    $n=4$ & $1.4\times 10^{-15}$ & $1.5\times 10^{-8}$ \\ 
    
    $n=10$& $1.8 \times 10^{-14}$ & $8.6\times 10^{-7}$   
    
  \end{tabular}
  }
  \caption{The minimum and maximum axion mass which can be probed by the 21cm forest for $n=0,4,10$.}
  \label{table:mass}
\end{table}

\section{Discussion/Conclusion}

\begin{figure}[htbp]
\includegraphics[width=1.0\hsize]{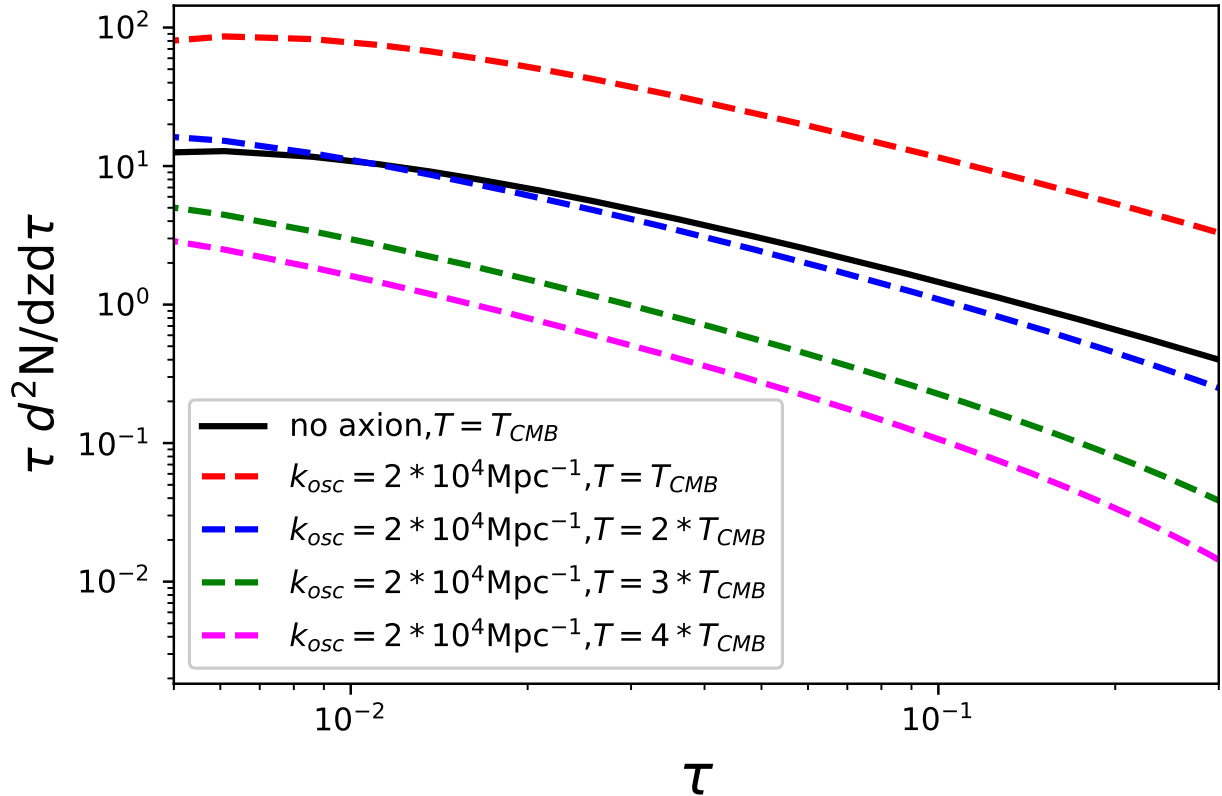}

\caption{
The illustration for the number of 21cm absorption lines with different gas temperatures.}
\label{fig:21cm_T}
\end{figure}

We first discuss the effect of the thermal evolution of the IGM. Our analysis so far employed the adiabatic cooling of the IGM. However, this assumption would be optimistic because of ignoring the highly energetic X-ray photons which can potentially heat up the IGM. The axion isocurvature fluctuations can enhance the minihalo abundance and also generate more X-ray heating sources. The heated IGM increases the Jeans mass and leads to the suppression of the 21cm absorption line abundance. Such a suppression is illustrated in Fig. \ref{fig:21cm_T} which shows how the 21cm forest observations would be affected by the different gas temperature for $k_{{\rm osc}}=2\times 10^4$/Mpc with $T_{{\rm IGM}}=T_{{\rm CMB}}\sim 4 \times T_{{\rm CMB}}$. The scenario with a relatively low $T_{{\rm IGM}}=T_{{\rm CMB}}=30$[K] at $z=10$ is still allowed by the recent observation \citep{2015ApJ...809...62P}. Even with the suppression due to a relatively high IGM temperature, we can still find that the number of 21cm absorption lines remains $O(1)$ at $\tau \sim 0.01$ up to $T= 4\times T_{{\rm CMB}}$. It  becomes less than unity at $\tau \gtrsim 0.01$ because the minihalos that have a large optical depth are removed in our calculations due to the increased Jeans mass ($\tau \propto T_{vir}^{-1}$). This result shows that we can still explore $k_{{\rm osc}}$ up to $2 \times 10^4 {{\rm Mpc}}^{-1}$ if the IGM temperature is less than $T= 4\times T_{{\rm CMB}}$. 
We also point out the possible degeneracies between the X-ray heating efficiency parameters (see,e.g. \cite{2013JCAP...09..014C}) and the axion parameters, which can be seen from the analogous behaviors of the 21cm forest signals in Fig. \ref{fig:21cm_T} with different gas temperatures and those in Fig.\ref{fig:21cm_abs} with different axion parameters.

The advantage of the 21cm forest observations is that it uses the 21cm absorption spectra from bright sources and it does not suffer from diffuse foregrounds which are challenging obstacles in the 21cm emission lines. The disadvantage is, on the other hand, that the 21cm forest relies on the existence of radio bright sources at a high redshift. According to \cite{2006MNRAS.370.1867F}, the required minimum brightness of a radio background source for the 21cm forest observation is given by

\begin{equation}
\begin{split}
S_{\rm min}=10.4\rm{mJy}\left(\frac{0.01}{\tau}\right)\left(\frac{S/N}{5}\right)\left(\frac{\rm{1 kHz}}{\Delta \nu}\right)^{1/2} \left(\frac{5000[\rm m^{2}/K]}{A_{\rm{eff}}/T_{\rm{sys}}}\right)\left(\frac{100~\rm{hr}}{t_{{\rm int}}}\right)^{1/2},
\end{split}
\label{eq:Smin}
\end{equation}
where $\tau$ is the target 21cm optical depth, $\Delta \nu$ is a frequency resolution, $A_{{\rm eff}}/T_{{\rm sys}}$ is the ratio between an effective collecting area and a system temperature and $t_{{\rm int}}$ is the observation time. In eq.(\ref{eq:Smin}), we normalise each quantity by the SKA-like specifications.

Recently, some radio bright sources, such as radio loud quasars (around 10\% of quasars are estimated to be radio emission dominant) and Gamma-ray burst (GRB), have been found \cite{2020A&A...635L...7B,ban2018,ban2015,2011ApJ...736....7C}. For example, the promising findings of the radio loud quasars and blazars with a sufficient brightness $\gtrsim {\cal O}(10)$ mJy have been reported at a redshift $z\gtrsim 6$ \cite{ban2015,ban2018, 2020A&A...635L...7B}. The estimates of the number of radio quasars based on the extrapolations of the observed radio luminosity functions to the higher redshift indicate that there could be as many as $\sim 10^{4}-10^{5}$ radio loud quasars with a sufficient brightness at $z$=10\cite{Haiman_2004,Xu_2009,Ivezic:2002gh}. Thus, these observations and prediction support the 21cm forest studies. We should mention that the existence of the relatively cold IGM is also preferable for 21cm forest observations in addition to a large number of bright radio sources at a high redshift. The thermal state of the IGM is, however, still uncertain and the current data only provides a loose constraint requiring $T_{{\rm IGM}}\gtrsim 7$ K at $z=8.1$ \citep{2015ApJ...809...62P}. If the 21cm forest detection turns out to be challenging due to the high $T_{{\rm IGM}}$ at a high redshift, we could obtain the tight bounds on the early X-ray heating scenarios.

We also mention the axion dark matter Jeans scale due to the quantum pressure inside which the pressure support prevents the matter fluctuation growth 
$k_\mathrm{J}(R)
     =\left(16\pi G R \rho_{a}(T_{\mathrm{now}})  \right)^{1/4} m_{a,0}^{1/2}$\cite{sv2006,Arvanitaki:2009fg, hu2000,ame2005,Hui:2016ltb,marsh15,Kadota:2013iya,sch2018,2017PhRvL.119c1302I}. 
Such a Jeans scale is quite small $k_\mathrm{J}(R_{{\rm eq}})>k_{\mathrm{osc}}$ for the parameter range of our interest and does not affect our discussions (i.e. the mass enclosed inside the Jeans scale is smaller than that inside $k_{\mathrm{osc}}$ scale when the dark matter overdense regions started collapsing to from the halos around the radiation-matter equality). Such a small scale suppression however can well be measurable by the future 21cm observations for the scenarios different from the ones discussed in this paper. For instance, Ref. \cite{shi2019} showed that the 21cm forest observations can probe such small scale suppression through the decrease of the 21cm absorption lines for the axion mass up to the order of ${\cal O}(10^{-18})$ eV (such studies apply for the scenarios, for instance, where the PQ symmetry breaks during the inflation and is never restored afterwards ). It is interesting to find that post-inflation scenarios discussed in this paper can lead to the different signature from such previous studies in that the enhancement, rather than the suppression, of the 21cm forest absorption lines can occur and also the sensitive axion mass range is bigger ($m_a\gtrsim {\cal O}( 10^{-18})$ eV) than those for pre-inflation PQ breaking scenarios.

While we discussed the 21cm absorption signals from the minihalos, the 21cm emissions from the minihalos can also give the complimentary probes on the axion dark matter which will be discussed in the forthcoming paper.

%
%
 %




\begin{acknowledgments}
 We thank the anonymous referee for the constructive comments on our manuscript. KK is supported by the Institute for Basic Science (IBS-R018-D1). KI is supported in part by JSPS KAKENHI Grant Numbers 18K03616 and 17H01110.
\end{acknowledgments}


\bibliography{ref}


\end{document}